\documentclass[letterpaper,pdftex,rmp,twocolumn,amsmath,amssymb,groupedaddress,floatfix]{revtex4}

\usepackage{graphicx}
\usepackage{natbib}
\usepackage{textcomp}
\usepackage[normalem]{ulem}
\usepackage{makecell}
\usepackage{float}
\usepackage{setspace}
\usepackage[a]{esvect}
\usepackage[font=sf,size=footnotesize,justification=RaggedRight,singlelinecheck=false]{caption}
\usepackage[sf,bf,small]{titlesec}
\titlespacing*{\section}{0pt}{1em}{0em}
\usepackage{xcolor}
\usepackage{placeins}

\usepackage{mathrsfs}
\usepackage{amsmath,amssymb}
\usepackage{bm}% bold math

\usepackage{float}

\definecolor{darkgray}{rgb}{0.25,0.25,0.25}
\definecolor{darkred}{rgb}{0.89,0.10,0.11}
\definecolor{darkblue}{rgb}{0.12,0.39,0.62}
\usepackage{url}

\usepackage[pdftex,breaklinks=true,colorlinks=true,citecolor=black,linkcolor=black,menucolor=black,urlcolor=darkblue,pdfborder={1 0 0}]{hyperref}
\hypersetup{pdftitle={............},pdfauthor={L.E.C. Rocha (2016)}}
\bibpunct{()}{)}{,}{s}{,}{,}

\usepackage{multirow}
\usepackage{booktabs}
\usepackage{array}
\usepackage{color,soul}

\begin{document}
\makeatletter
\renewcommand\@biblabel[1]{#1.}
\makeatother

\newcommand{\dif}{\mathrm{d}}

\renewcommand{\figurename}{Figure}
\renewcommand{\thefigure}{\arabic{figure}}
\renewcommand{\tablename}{Table}
\renewcommand{\thetable}{\arabic{table}}
\renewcommand{\refname}{\large References}

\addtolength{\textheight}{1cm}
\addtolength{\textwidth}{1cm}
\addtolength{\hoffset}{-0.5cm}

\setlength{\belowcaptionskip}{1ex}
\setlength{\textfloatsep}{2ex}
\setlength{\dbltextfloatsep}{2ex}

\title{Modeling contact networks of patients and MRSA spread in Swedish hospitals}

\date{\today}

\author{Luis E C Rocha}
\email{luis.rocha@ki.se}
\affiliation{
Department of Public Health Sciences, Karolinska Institutet, Stockholm, Sweden \\
Department of Mathematics and naXys, Universit\'e de Namur, Namur, Belgium}

\author{Vikramjit Singh‎}
\affiliation{Universit\"at Bonn, Bonn, Germany}

\author{Markus Esch}
\affiliation{Fraunhofer Institute for Communication, Information Processing and Ergonomics, Wachtberg, Germany}

\author{Tom Lenaerts}
\affiliation{Universit\'e Libre de Bruxelles, Brussels, Belgium}

\author{Mikael Stenhem}
\affiliation{S\"ormland County Health Council, Sweden}

\author{Fredrik Liljeros}
\affiliation{Department of Sociology, Stockholm Universitet, Stockholm, Sweden}

\author{Anna Thorson}
\affiliation{Department of Public Health Sciences, Karolinska Institutet, Stockholm, Sweden}

\begin{abstract}
Methicillin-resistant Staphylococcus aureus (MRSA) is a difficult-to-treat infection that only in the European Union affects about 150,000 patients and causes extra costs of 380 million Euros annually to the health-care systems. Increasing efforts have been taken to mitigate the epidemics and to avoid potential outbreaks in low endemic settings. Understanding the population dynamics of MRSA through modeling is essential to identify the causal mechanisms driving the epidemics and to generalize conclusions to different contexts. We develop an innovative high-resolution spatiotemporal contact network model of interactions between patients to reproduce the hospital population in the context of the Stockholm County in Sweden and simulate the spread of MRSA within this population. Our model captures the spatial and temporal heterogeneities caused by human behavior and by the dynamics of mobility within wards and hospitals. We estimate that in this population the epidemic threshold is at about 0.008. We also identify that these heterogeneous contact patterns cause the emergence of super-spreader patients and a polynomial growth of the epidemic curve. We finaly study the effect of standard intervention control strategies and identify that screening is more effective than improved hygienics in order to cause smaller or null outbreaks.
\end{abstract}

\maketitle

\noindent 

The increasing tolerance of some bacteria to currently available antibiotics has become a major public health issue in recent decades~\cite{ECDC2009, WHO2014}. Hospital-acquired (HA) Methicillin-resistant Staphylococcus aureus (MRSA) infection has been routinely detected in hospitalized patients including those in high-income countries. In the European Union 150,000 patients are affected annually~\cite{Kock2010}. In Sweden, for instance, the incidence rate (per 100,000 people) of MRSA jumped from 10.76 in 2005 to 29.96 in 2014~\cite{PHAS2015}. Although some strains may be harmless to healthy people or are not health-care associated, such difficult-to-treat infections are particularly dangerous in a context of individuals with weakened immune systems~\cite{vanHal2012, Hanberger2011}. Bacteria resistance may develop spontaneously, but commonly, a healthy person become infected via direct contact with an infected host, or contaminated devices and surfaces~\cite{CNUK2005}. A hospital setting, if not under strict hygienic control, provides excellent conditions for efficient spread of MRSA. Although colonization typically occurs in the anterior nares, open wounds, urinary infections, or intravenous catheters are also potential sites for infection. Therefore, the daily contact between health care workers (HCW) and patients is sufficient for the propagation of the pathogens. This is worsened because colonized individuals, even if not ill, may still infect others. During a regular shift, HCWs typically interact with various patients whereas patients usually interact with different HCWs. These interactions create dynamic contact networks~\cite{Bansal2010, Masuda2013} in which the MRSA infection eventually propagates. These contact networks have a complex structure of who was in contact with whom at a given time because of the non-trivial dynamics of a typical day in a hospital~\cite{Vanhems2013}. Although the transmission between a patient and a HCW may be more likely in certain wards (e.g. burns or transplant unit~\cite{CNUK2005}), the mobility of patients between wards or hospitals creates the missing links responsible to sustain the spread of HA-MRSA. Therefore, identifying which contact patterns (or network structures~\cite{Newman2010}) regulate the propagation of the infection is the first step to better understand the spread potential of MRSA, and then to develop efficient protocols to reduce the incidence in endemic areas and to avoid potential outbreaks in low-prevalence contexts~\cite{Stenhem2006}.

\begin{figure*}[htb]
\centering
\includegraphics[scale=1]{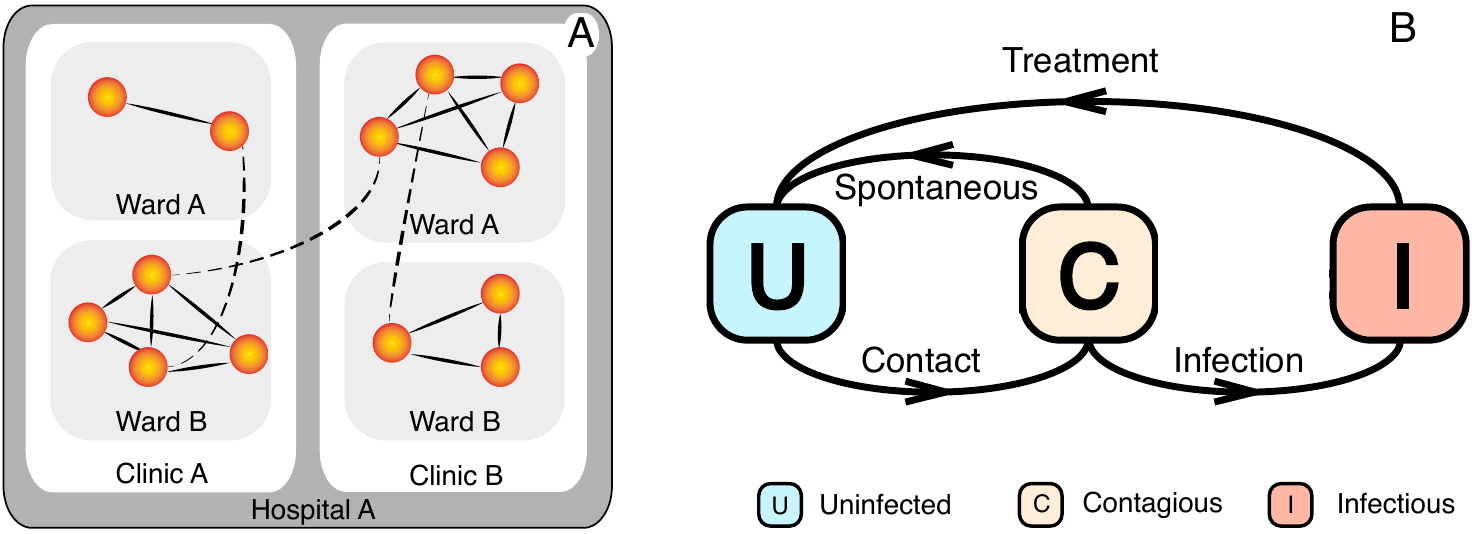}
\caption{\textbf{Architecture of the population and the MRSA contagion model.} (a) shows the structure of the contact network within the various levels of the hospital system and (b) shows the potential transitions between states in the MRSA infectious disease model.}
\label{fig01}
\end{figure*}

Mechanistic models aim to generate an abstract mathematical representation of the most relevant characteristics of a system. The gain in understanding, particularly of the causal relations, by simplifying the problem compensates the information that is discarded. This approach, successfully used in other disciplines to e.g. forecast the weather or the movement of planets, fundamentally differs from traditional epidemiological methods such as controlled experiments and logistic regression models. A mechanistic model allows assessing realistic scenarios without the need to experiment on the real population. These models in fact have been used for a long time to study the spread of infections at the population level~\cite{Diekmann2012, Kelling2007}. In the context of MRSA, a number of studies have been published in recent years mostly using the frameworks of compartmental and agent-based models~\cite{Kleef2013}. Compartmental models are convenient because they can be elegantly described by a set of coupled differential equations, used to describe a group of well-mixed individuals in a certain state~\cite{Cooper2004, Bootsma2006}, and can be sometimes studied analytically~\cite{Bootsma2006,Simon2013}. Agent-based models, on the other hand, request more computational resources but are more flexible and allow the addition of different levels of complexity by defining updating and interaction rules for each individual~\cite{Macal2011, Lee2011}. Previous models typically focus either on the population dynamics within a single ward~\cite{Hall2012} or in a single hospital with a few wards~\cite{Bootsma2006,Sadsad2013}, between one hospital and the community~\cite{Sadsad2013}, or between multiple hospitals with a simple ward structure~\cite{Bootsma2006,Lee2011}. As in any modeling exercise, these studies make a series of assumptions regarding different aspects of the population. In particular, exponential probability distributions are defined to account for the patient’s length-of-stay, readmission, or referral to another hospital~\cite{Cooper2004, Bootsma2006, Simon2013}. Even in models with some structure, i.e. those that have more than one ward or one hospital, average values are used to characterize similar units, as for example, the size of the hospital or the frequency of interactions between HCWs and patients in each ward~\cite{Bootsma2006,Lee2011}. In other words, these previous models attempted to reproduce the hospital population but failed to fully capture the heterogeneities and complexity of the contact patterns and patient mobility, emerging as a consequence of referral and readmission of patients within and between hospitals.

In this paper, we propose a network-based model of the full hospitalized patients of the Stockholm county in Sweden, a population of about 2,192,433 inhabitants. By using information of patient flow (inpatients), we are able to reconstruct, at the individual level, the potential contacts between all patients in all hospitals and care units of the county. These contacts, mediated by HCW or contamined objects and surfaces, are the most likely pathways for the spread of HA-MRSA. We assume that a contact exists between two patients if they have been hospitalized in the same ward at the same time. This methodology naturally captures realistic contact patterns and avoids several assumptions regarding the dynamics of patients as for example, length-of-stay, readmitance, mobility between wards and hospitals, hospital size, etc. Using realistic information is particularly important since people are different and contact patterns cannot be characterized by the average behavior. We show that the realistic contact patterns observed in this dataset are responsible for the appearence of few highly contagious individuals and for the polynomial growth of the prevalence curve in the absence of community infected individuals. Furthermore, we observe that the probability distribution of the final outbreak is quite heterogeneous with high likelihood of small or large outbreaks. We also study the effect of screening upon hospital admission and improved hygienics on the outbreak size after one year of the onset of the epidemics.

\section*{Model}

\noindent\textbf{Patient Flow Data Set.} We gather information on the admission and discharge dates of 743,182 patients in 485 hospitals and nursing homes at 52 different geographical locations in the Stockholm County, Sweden. This information is collect during 3,059 continuous days in the 2000s (the exact years are confidential for ethical reasons). A total of 2,019,236 admissions are recorded. Each patient and the respective ward of hospitalization are anonymous but both have unique IDs for identification.

\noindent\textbf{Contact Network Model.} A contact network is a set of links connecting pairs of vertices~\cite{Bansal2010, Masuda2013}. Our contact network model is formed by patients (the vertices) that shared a ward at the same time (Fig.~\ref{fig01}A). We assume that patient-to-patient contacts occurs through hidden HCWs or contaminated objects and surfaces. We discard contacts made in the community (non-hospitalized population). The time resolution is one day, implying that the contact network changes in time, i.e. a link (representing the contact) between two individuals may exist or not at a given time~\cite{Bansal2010, Masuda2013}. After each day, links may appear, disappear or be maintained according to real-life patient's dynamics.

\vspace{0.5cm}
\noindent\textbf{Contagion Model.} We assume that a patient may be either Uninfected (U), Contagious (C), or Infected (I)~\cite{Kajita2007}. Transitions between these states occur according to the interaction dynamics or according to the progression of the infection (Fig.~\ref{fig01}B). An uninfected patient can be infected upon contact, with probability $\beta_{\text{C}}$ and $\beta_{\text{I}}$, from a patient that is C or I, respectively. An individual in the states C or I cannot be re-infected. Upon infection, i.e. when turning to state C, the person will develop the infection with probability $\mu = 0.2$~\cite{Kajita2007} or naturally recover with $1-\mu$. If the person develops the infection, it moves from state C to I after $\tau_{\text{infec}} = 9.5$ days~\cite{Kajita2007, Sadsad2013}. On the other hand, if the person naturally recovers, it moves from state C to U after an average of $\tau_{\text{rec}} = 370$ days~\cite{Cooper2004}. We assume that an admitted patient may be contagious with probability $\alpha_{\text{adm}}$~\cite{Hall2012} and consequently uninfected with $1-\alpha_{\text{adm}}$. If treatment occurs, the patient is cured after $\tau_{\text{treat}}$ and then can be reinfected. Table~\ref{tab01} summarizes the parameters of the model. Discharged patients are assumed to be uninfected. We also assume antibiotic resistance does not emerge spontaneously in this population during the study period.

\begin{table}[h]
    \begin{tabular}{ll}
    Parameter                                                    & Value                                                   \\ \hline  
    Per-contact Infection probability if C            & $\beta_{\text{C}}$                                \\
    Per-contact Infection probability if I              & $\beta_{\text{I}}$                                 \\
    Probability to develop infection                    & $\mu=0.2$                                           \\
    Prob. admitted infected                               & $\alpha_{\text{adm}}$                          \\
    Time at stage C if developing infec.            & $\tau_{\text{infec}}=$ 9.5 days             \\
    Time at stage C if not developing infec.      & $\tau_{\text{rec}} = 370$ days             \\
    Time to cure if treated                                 & $\tau_{\text{treat}} =$ 7 days               \\ \hline
    \end{tabular}
\caption{\textbf{Parameters of the model.} The table shows the parameters and respective values, if fixed, of the model.}  
\label{tab01}
\end{table}

\section*{Results}

\subsection*{Patient Dynamics}

\begin{figure}[htb]
\centering
\includegraphics[scale=0.9]{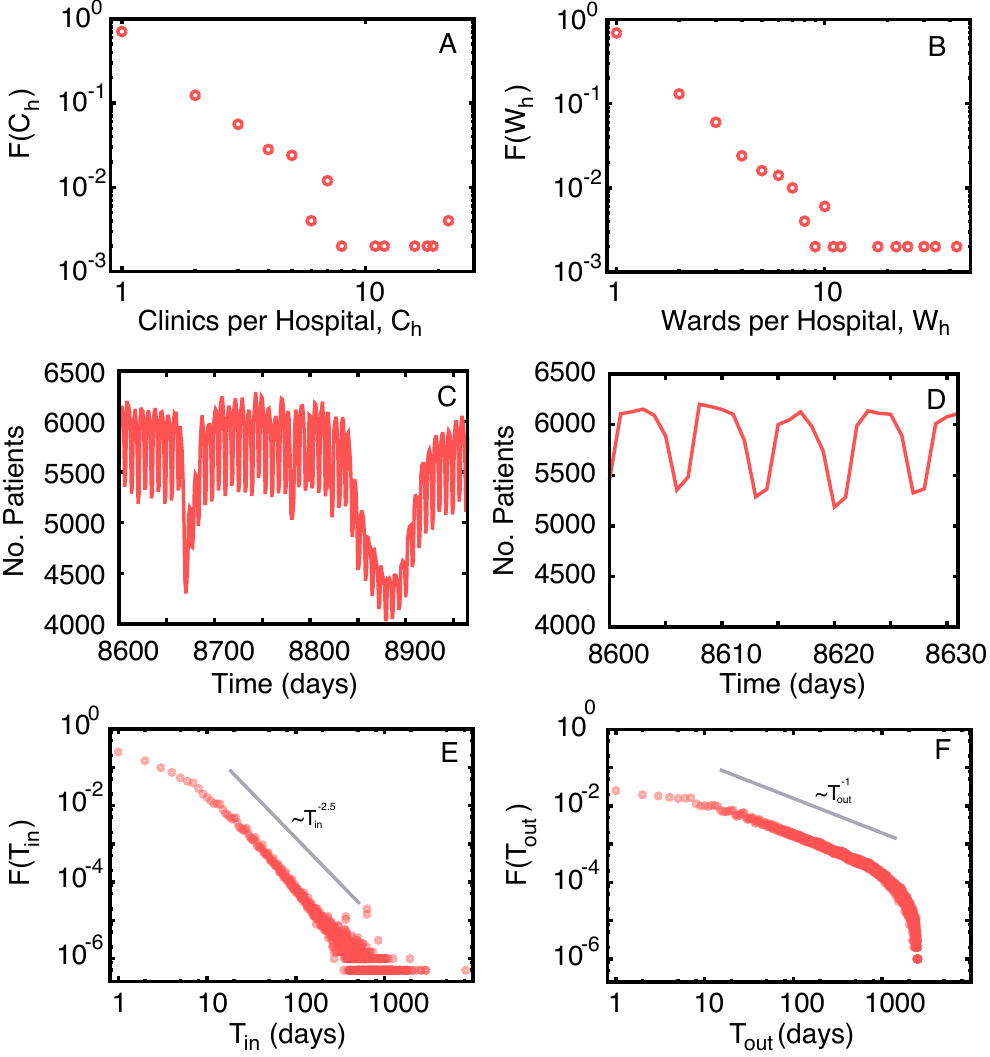}
\caption{\textbf{Spatial and temporal patterns.} Fraction of hospitals with a given number of (A) clinics ($C_h$) or (B) wards ($W_h$), both axes are in log-scale; Typical (C) annual and (D) weekly cycles; Distribution (E) of the duration of hospital stays and (F) of the times between two subsequent hospitalizations by the same patient. Both axes are in log-scale.}
\label{fig02}
\end{figure}

The hospital system contains 859 clinics and 979 wards distributed within 485 hospitals. There is an assymetry in which the majority of the hospitals and health care units contain a single clinic and a single ward whereas a few rarer hospitals and health care units contain more than 10 clinics and wards (Fig. \ref{fig02}A,B). Although this observation may be expected from real-life experience, this diversity of hospital structure has important consequences on modeling since it shows that hospitals are different in terms of internal structure and thus uniform or homogenous assumptions inneficiently represent the hospital system. The dynamics of hospitalizations also follows non-trivial patterns. There are strong weekly and annual patterns in which the hospital population may decrease about $25\%$ (Fig. \ref{fig02}C) during summer and winter breaks, and about $10\%$ (Fig. \ref{fig02}D) over the weekend in comparison to week-days.

The duration of hospitalizations ($T_{\text{in}}$) and times between subsequent hospitalizations ($T_{\text{out}}$) also follow non-trivial patterns. Figure~\ref{fig02}E,F shows that the distribution of both measures strongly differ from the typical homogeneous (exponential) assumption. In fact, the vast majority of patients spend a few days in the hospital whereas a substantially reduced number of patients may spend over a year in there~\cite{Liljeros2007}. Long stays both increase the risk of infection and of further spread in case the respective patient becomes infected. Similarly, the interval between two hospitalizations shows that most patients are readmitted within few weeks after discharge whereas few of them return to the hospital more than a year after being discharged. In other words, these results mean that hospitalizations tend to be short but frequent, a dynamics that may contribute to community infection if discharged patients go home contagious and make frequent contacts with other individuals. We also observe that the probability to return to the hospital (re-admission) decreases with the time between hospitalizations (see supplementary Figure~\ref{fig0X}A). Similarly, the average hospitalization times decrease with the time between two hospitalizations (see supplementary Figure~\ref{fig0X}B). These temporal correlations are typically discarded when modeling the population dynamics of MRSA. The consequence of this heterogeneity is that individuals are not the same and thus have different relevance in the spread of an infection.

\subsection*{Infection Dynamics}

\begin{figure}[htb]
\centering
\includegraphics[scale=1.3]{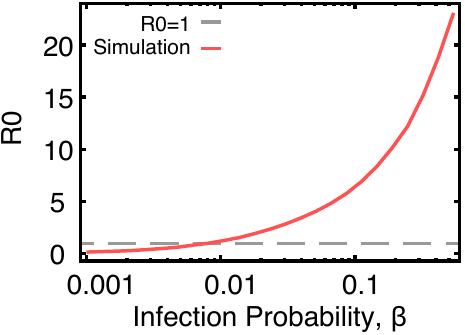}
\caption{\textbf{Reproduction number.} The relation between the infection probability $\beta$ and the reproduction number (secondary infections) according to simulations of the spread of the infection. The dashed line at $R0=1$ represents the epidemic threshold. The x-axis is in log-scale.}
\label{fig03}
\end{figure}

To study the population dynamics of simulated HA-MRSA in the Swedish context, we select one year of the original data set and extract the contact patterns between the patients. This sample has 170,839 patients and 20,483,587 contacts. For simplicity, we assume that $\beta_{\text{C}} = \beta_{\text{I}} = \beta$ and scan the values of the per-contact infection probability $\beta$ to estimate the reproduction number $R0$ through simulations. We also assume $\alpha_{\text{adm}}=0$ and no treatment. $R0$ defines a threshold in which a large epidemic outbreak may occur if $R0 > 1$ or may not occur if $R0 < 1$. To estimate $R0$, we first infect a single individual at time 0 (zero) and set all other individuals as uninfected, then we simulate the spread of the infection following the infection dynamics (see section "Contagion Model") and count the number of secondary infections made by the seed infected individual. We repeat this procedure 20000 times to calculate the statistics. Figure~\ref{fig03} shows that the epidemic threshold is at $\beta\sim0.008$, therefore implying that if the infection probability is larger than $\beta\sim0.008$, a large epidemic outbreak is likely to happen in a fully uninfected population.

\begin{figure}[htb]
\centering
\includegraphics[scale=0.9]{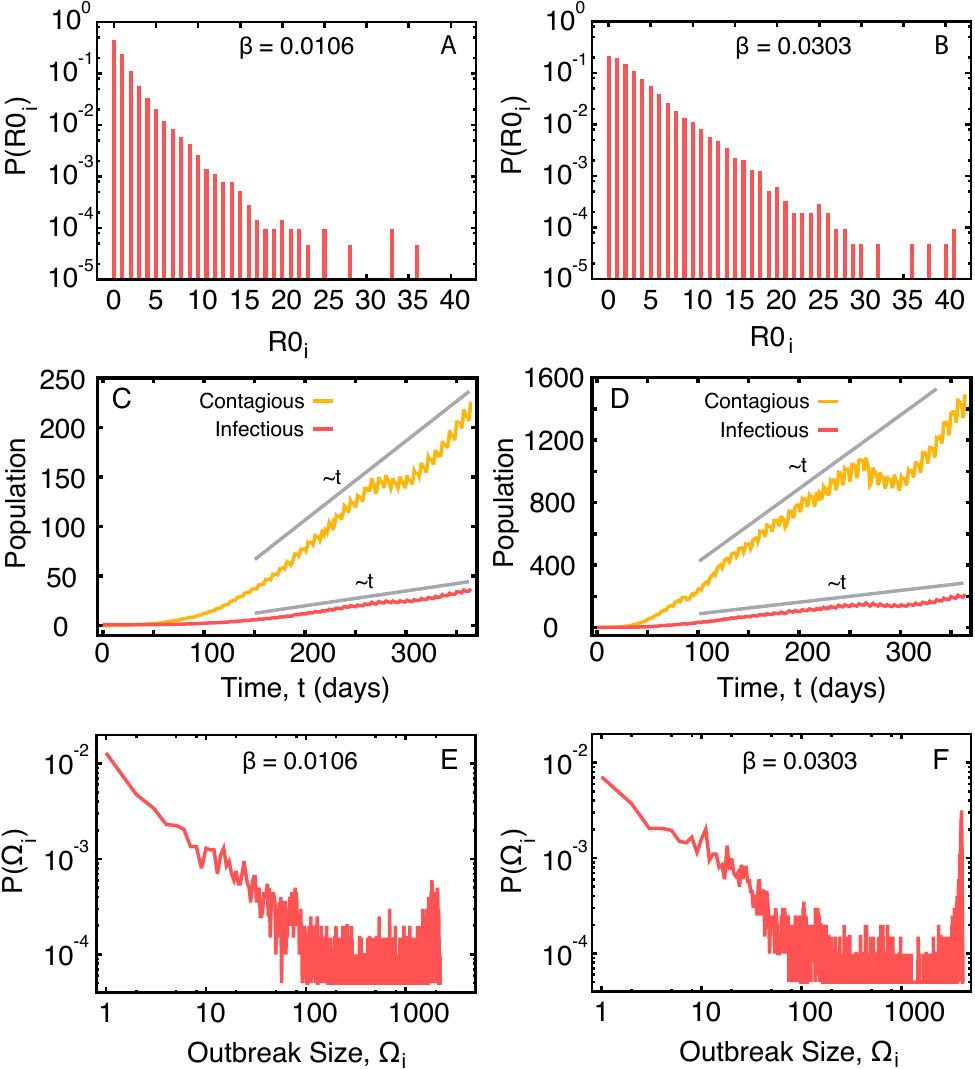}
\caption{\textbf{Evolution of the state of the patients.} The reproduction number for (A) low ($\beta = 0.0106$) and (B) high ($\beta = 0.0303$) per-contact infection probability. y-axes are in log-scale. The evolution of the number of contagious and infectious individuals for (C) low ($\beta = 0.0106$) and (D) high ($\beta = 0.0303$) per-contact infection probabilities. The final outbreak size (at $T = 365$ days) for (E) low ($\beta = 0.0106$) and (F) high ($\beta = 0.0303$) per-contact infection probabilities. Both axes are in log-scale.}
\label{fig04}
\end{figure}

The heterogeneity of the network structure implies that patients do not have the same role. Looking at the distribution of R0, i.e. the number of secondary infections when we start the infection at random patients, we observe that, for both low and high per-contact infection probability, most nodes cause zero or few secondary infections whereas a very few particular patients may cause up to 40 secondary infections (Fig.~\ref{fig04}A,B), which may characterize them as super-spreaders. This is a consequence of making too many contacts, possibly because they spend a much larger than average time in the hospital.

We investigate the growth curve of the epidemic outbreak for two values of the per-contact infection probability, $\beta=0.0106$ (low) and $\beta=0.0303$ (high). Figure~\ref{fig04}C,D shows the average number of infectious (red curve) and contagious (yellow curve) individuals in time for both infection probabilities. It takes around 100 days for the number of infectious individuals to take off, yet, in both cases, the curves are characterized by a (trend) polynomial linear growth after the initial non-linear regime (Fig.~\ref{fig04}C,D). The effect of the lower number of patients during holidays appears at around 200 days after the onset of the epidemics (the location of this decrease depends on the starting dates of the simulation), however, the decrease in the number of contagious and infectious individuals is not as significant as the decrease on the total number of patients. It is also possible to see that the weekly pattern reflects on the number of contagious individuals and to a lesser extent on the number of infectious individuals. The distribution of outbreak sizes $\Omega_{\text{i}}$ (contagious plus infectious individuals) also has a particular bimodal shape. It indicates that a large number of outbreaks will cause less than 100 infections but on the other hand there is a non-negligible chance of large outbreaks affecting more than 1000 individuals.

\subsection*{Infection control}

We study the effect of screening and hygienics in the infection dynamics. To do so, we assume that newly admitted patients may be contagious with a probability $\alpha_{\text{adm}} = 0.0877$~\cite{Hall2012}, implying on a constant influx of contagious patients that is different from the assumptions of the previous section. Furthermore, we assume that screened and infectious patients automatically undergo treatment, lasting on average $\tau_{\text{treat}} = 7$ days. During treatment, patients are quarantined and cannot infect or be re-infected.

\begin{figure}[htb]
\centering
\includegraphics[scale=0.9]{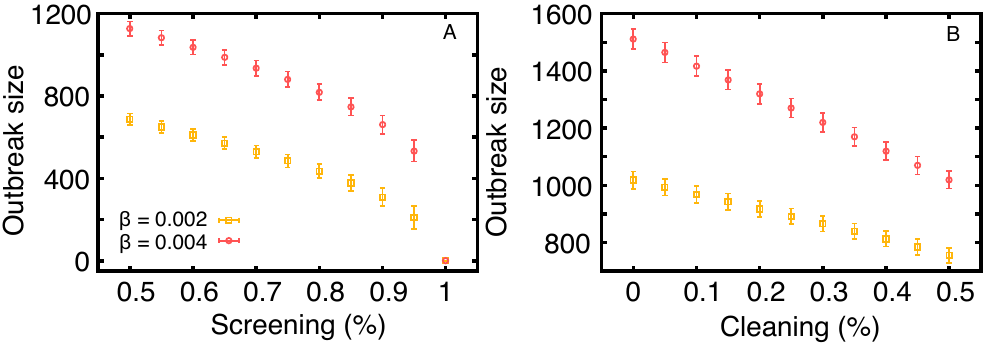}
\caption{\textbf{Screening and improved hygienics.} (A) the outbreak size vs. increased hygienics for low ($\beta = 0.002$) and high ($\beta = 0.004$) per-contact infection probability; (B) the outbreak size vs. screening on admitance for (a) low ($\beta = 0.002$) and high ($\beta = 0.004$) per-contact infection probabilities. In both cases, the final outbreak size $\Omega$ is calculated as the average outbreak on the last week of the data, i.e. $T=[359,365]$.}
\label{fig05}
\end{figure}

The constant influx of multiple contagious patients implies that a lower per-contact infection probability is able to sustain the epidemics. We thus test two scenarios, with $\beta=0.002$ (low) and $\beta=0.004$ (high), i.e. bellow the epidemic threshold. To simulate screening, we select a random fraction $\psi$ of admitted patients and put them into treatment if contagious; uninfected patients are released. Figure~\ref{fig05}A shows that for both infection probabilities, increasing the fraction of screened patients decreases the final outbreak size $\Omega$ (we estimate $\Omega$ by taking the average number of contagious and infectious patients in the last week of the data, i.e. $T=[359,365]$). However, irrespective of $\beta$, full screening is necessary to reduce $\Omega$ to zero. To simulate increased hygienics, for example by properly washing hands before touching patients, we decrease the per-contact infection probability by a factor $1-\phi$, that is, we set the per-contact infection probability as $\beta*(1-\phi)$. Figure~\ref{fig05}B shows that improving by $50\%$ the hygienics is insufficient to reduce by half the outbreak size (measured in the same way as for the screening). Relatively speaking, screening turns out to be more effective than improving hygienics in this population.

\section*{\large Discussion}

Antibiotic-resistant bacteria has been increasingly pressing hospital systems in both low and high-income countries. In particular, Methicillin-resistant Staphylococcus aureus (MRSA) is a difficult to treat bacteria that can be particularly dangerous to people with weakened immune system such as hospitalized individuals. Although antibiotic-resistance has developed due to intense use of antibiotics, MRSA commonly spreads through physical contacts with infected individuals, objects or surfaces. In order to understand this dynamics of infections, we develop a data-driven model of contact patterns between patients in a large hospital system in Sweden. Such model is able to capture spatial and temporal heterogeneities, inherent to hospital structures and patients behavior, into the network structure.

We simulate the spread of MRSA in a fully uninfected population and estimate that the epidemic threshold ($R0 = 1$) is relatively high, that is, it occurs at $\beta \sim 0.008$. Nevertheless, this threshold is lower in case of a constant influx of contagious individuals, as typically occur in real systems. We observe that right above the threshold, the epidemic curve grows linearly (after an initial shorter non-linear growth) as a consequence of the heterogeneities in the contact patterns that constrain the spread and avoid an exponential growth, typically observed in theoretical models. These heterogeneous contact patterns also cause the appearance of a few super-spreaders, that is, individuals infecting a much larger than average number of patients. Another consequence is the bimodal distribution of outbreak sizes in which the probability of minor outbreaks ($< 100$ infected individuals) is relatively high, but larger outbreaks ($> 1000$ individuals) are not uncommon. This is in accordance to previous reports of MRSA outbreaks in Sweden during the 1990s~\cite{Stenhem2006}.

The goal of modeling exercises is to understand the main mechanisms sustaining the spread of MRSA in order to reduce its incidence and hence mortality and costs. The analysis of the population dynamics of the infection is often done by comparing the implementation of some infection control protocol against the absence of any action. It is generally believed that hand hygiene is an effective means to avoid the propagation of MRSA~\cite{WHO2014,CNUK2005}; the challenge however lies in enforcing such routine in daily life. We show that reducing infection, as a consequence of improved hygienics, by $50\%$ is not sufficient to half the final outbreak size. Since sufficient cleaning of hands and utensils is many times not achievable, other strategies involving screening followed by isolation of colonized or infected people have to be introduced. As we have shown, screening every admitted patient is a priori ideal but as previously reported~\cite{Gurieva2013}, associated to high financial costs. Screening of intensive care patients or previously documented carriers, on the other hand, has been suggested as cost-effective alternatives to global screening~\cite{Gurieva2013}. Since hospitals are strongly connected through transfer of patients, a surveillance system based on a few sentinel hospitals, chosen according to their centrality in this referral network, may be also an effective means for early detection, and thus control, of MRSA outbreaks~\cite{Ciccolini2013}. An agreement on the best cost-effective policies however is still missing.

%\section*{\large Acknowledgements} 

%\section*{\large Author Contributions}

\section*{\large Ethics Statement}The Regional Ethical Review Board in Stockholm approved the data about inpatients in Sweden (Record Number 2006/3:3). No informed consent was obtained but all data are anonymous. There is no information about the identity of individuals or other personal characteristics that could be linked to specific individuals. The only personal data that will be used are the admission and discharge dates of each anonymous patient.

\section*{\large Additional information} The authors declare that they have no competing financial interests. Correspondence and requests for materials should be addressed to LECR~\mbox{(luis.rocha@ki.se)}.

\bibliographystyle{unsrt}
\bibliography{MRSA_spread}

\end{document}